\documentclass{article}

\usepackage{amsmath,mathrsfs}
\usepackage{ upgreek }
\usepackage{ amssymb }
\usepackage{graphicx,subfigure}
\graphicspath{{v-phi/}}
\usepackage{multirow,booktabs}
\usepackage{makecell,rotating,multirow,diagbox}
\usepackage[numbers,sort&compress]{natbib}
\usepackage{indentfirst}

\def\abstract#1{\vskip 7mm 
        \begin{center}{\large Abstract}\par \smallskip
                \begin{minipage}[c]{12cm}
                        \small #1
                \end{minipage}
        \end{center}
}
\def\title#1{\begin{center}{\Large\bf #1}\end{center}}
\def\author#1{\vskip 5mm \begin{center}{#1}\end{center}}
\def\address#1{\begin{center}{\it #1}\end{center}}
\makeatletter
\@ifundefined{lesssim}{}{}
\@ifundefined{gtrsim}{}{}
\def\vereq#1#2{\lower3pt\vbox{\baselineskip1.5pt \lineskip1.5pt
\ialign{$\m@th#1\hfill##\hfil$\crcr#2\crcr\sim\crcr}}}
\makeatother

\begin{document}
\title{Dirac and Majorana neutrino scattering by cosmic torsion in spatial-flat FRW spacetime background}
\author{ Wei Lin$^{a,b}$, Xun Xue$^{a,b,c,}\footnote{Corresponding author:xxue@phy.ecnu.edu.cn}$}
\address{ $^a$Department of Physics, East China Normal University, Shanghai 200241, China}
\address{ $^b$Chongqing Institute of East China Normal University, Chongqing 401120, China}
\address{ $^c$Center for Theoretical Physics, Xinjiang University, Urumqi 830046, China}

\abstract{The possibility of distinguishing Dirac and Majorana fermions by cosmic torsion in the spatial-flat FRW spacetime is discussed. The scattering amplitudes of two types of fermions deviate from each other by the vector part of torsion in non-minimal coupling case. The scattering of massive fermions by cosmic torsion leads to a shift of final state energy distribution. The difference between shift values of two types of fermions can be used to distinguish fermion types of neutrinos.   }
\section{Introduction}

 The neutrino oscillation experiment reveals that neutrino has a tiny mass about 0.1eV\cite{Olive_2014}. However, the origin of the neutrino mass remains to be unclear. In the Standard Model, fermions get Dirac type masses via the Yukawa coupling by the Higgs mechanism. If neutrino has only the Dirac type of mass, the Yukawa coupling constants of neutrino would be smaller than ones of charged leptons at the order of $10^{-5}$, as the same order of their mass magnitudes difference, which is regarded as unnatural\cite{PhysRevLett.123.221802}. The see-saw mechanism can naturally explain the tiny mass of the left-hand neutrino in stead of assuming the huge difference between neutrino Yukawa coupling and charged lepton ones with the large Majorana mass for the right-hand neutrino. However the see-saw mechanism gives the Majorana type of masses to the left-hand neutrino which is the only observable part of neutrino at the Standard Model energy scale\cite{Yanagida:1979as}. Whether neutrino mass is a Dirac one or a Majorana one concerns different neutrino mass generation mechanism. The Majorana nature of neutrino mass provides a direct indication of the existence of new physics beyond Standard Model. Neutrinoless double beta decay process is a direct way to distinguish the Majorana type of neutrino from Dirac one. However, the experiment is not able to give a deterministic result to the existence of neutrinoless double beta decay yet\cite{Oberauer:2020mdv}. There are other proposals to determine the fermion type of neutrino, e.g., the idea to distinguish Dirac fermion from Majorana one by the scattering of fermion in the gravitational field\cite{Singh:2006ad,Nieves:2006tq,Singh:2006fn}. Lai and Xue consider the possibility to determine the fermion type of neutrino using its scattering behavior by torsion field\cite{Lai:2021tbw}.

Torsion can be decomposed into the vector part, the axial vector part and the pure tensor part according to the irreducible representation under the global Lorentz group\cite{Aldrovandi:2013wha}. In the minimal coupling scheme, only the axial vector torsion can couple with  spinor field. The vector torsion can have non-minimal coupling with spinor field while the pure tensor torsion is impossible to couple with spinor field even in the non-minimal coupling scheme\cite{Shapiro:1998zh,shapiro2002physical}. The non-minimal couple between torsion and spinor field is  universal due to the renormalization effect of spinor field\cite{BUCHBINDER1985263}.
\par The idea of distinguishing fermion types by torsion field is first proposed by Lai and Xue\cite{Lai:2021tbw} that the scattering by vector torsion field can distinguish the Dirac from Majorana neutrino in the non-minimal coupling case in the asymptotic Minkowski spacetime background. Although General Relativity is a torsion-free gravitation theory and has been verified by observations from solar system to galactic scale phenomenon\cite{Will:2005yc,Asmodelle:2017sxn,Crispino:2019yew}, the $H_0$ tension problem that the Hubble constant measured by cosmological model-independent standard candles calibration in the late-time universe and that calculated from CMB data based on $\Lambda CDM$ model has a discrepancy more than 3$\sigma$ in the recent result\cite{freedman2017cosmology} indicates that General relativity, the gravitation theory $\Lambda CDM$ based on, may need modifications from quantum gravity at cosmic scale\cite{DiValentino:2019qzk,DiValentino:2020hov}. Some proposals of modified cosmological models predict non-zero torsion distribution\cite{wu2016sim,Wu:2015jfa,yang2017effective,Wen-Ye:2017wfh,Li:2020tqx,Shen:2018elj,Zhai:2019std,Zhang:2017dvv} which will take effect on processes happened at cosmic scale, e.g., the cosmic neutrino propagation process. The effect of expanding universe must be taken into account during cosmic scale process so that the scattering of neutrino should be treated in a FRW background rather than a asymptotic Minkowski spacetime.

\section{Scattering amplitudes of fermions by torsion fields in spatial-flat FRW spacetime background}
\subsection{Interaction Hamiltonian density}
\par The scattering amplitude or S-matrix can be expressed as 
\begin{equation}\label{total_S-matrix}
 \textbf{S}=\textbf{T}\{e^{-i\int d^4x \mathcal{H}_{I}\left(x\right)}\}
\end{equation}
in QFT framework\cite{mandl2010quantum} where $\mathcal{H}_{I}$ is the interaction Hamiltonian. In the case of fermion coupling with background gravity, the interaction Hamiltonian can be read from the the Dirac action in the curved spacetime,
\begin{equation}\label{original_S_D}
 S_{D}=\int {{d^4}x} \sqrt { - g} \bar \psi \left[ \frac{i}{2} {\gamma^{\mu} }\overleftrightarrow{\mathcal{D}_{\mu}}\psi - m\psi \right]\;,
\end{equation}
where ${\mathcal{D}}_{\mu}=\partial_\mu-iA_{\mu}$ is the Fock-Ivanenko covariant derivative in minimally coupling scheme, $A_\mu=\dfrac{i}{2}A^{ab}_{\ \ \mu}S_{ab}$ is a Lorentz algebra valued 1-form known as the Lorentz connection or the spin connection and $S_{ab}$ are the Lorentz generators in a given representation\cite{Aldrovandi:2013wha}. By convention, spacetime indices is denoted by the lowercase Greek letters, e.g. $\mu,\nu,\rho,\cdots$ etc. while the tangent space indices by lowercase Latin letters, e.g. $a,b,c,\cdots$ etc. but Latin letters $i,j,k,\cdots$ are left for space indices of spacetime. Due to the equivalence principle, the tangent space can be equipped with local Lorentzian frame represented by tetrad fields $h_a=h_a{}^{\mu}\partial_\mu$ and the corresponding coframe $h^a=h^a{}_{\mu}dx^\mu$ satisfying $\displaystyle{g_{\mu\nu}=\eta_{ab}h^a_{\ \mu}h^b_{\ \nu}}$ and $\eta_{ab}=g_{\mu\nu}h_a^{\ \mu}h_b^{\ \nu}$ where $\eta_{ab}=diag\left(1,-1,-1,-1\right)$ is the  Minkowskian metric. For spinor field $\psi$, $S_{ab}$ are given by 
$\displaystyle{S_{ab}=\frac{i}{4}\left[\gamma_a,\gamma_b\right]}$ with $\gamma_a$ the Dirac matrices. The Lorentz connection can be decomposed into $\displaystyle{A^{ab}_{\ \ \mu}=\tilde A^{ab}_{\ \ \mu}+K^{ab}_{\ \ \mu}}$ where the contortion tensor $K^{ab}_{\ \ \mu}$ is defined by $\displaystyle{K_{abc}=\frac{1}{2}\left(T_{bac}+T_{cab}-T_{abc}\right)}$ and $\tilde A^{ab}_{\ \ \mu}$ is the torsionless Levi-Civita spin connection which is determined completely by the choice of local tetrad fields expressed as $\displaystyle{\tilde A_{abc}=\frac{1}{2}\left(f_{bac}+f_{cab}-f_{abc}\right)}$ and determines the curvature of spacetime completely in General Relativity, where the torsion fields $\displaystyle{T^{\rho}{}_{\nu\mu}}$ are defined by $\displaystyle{T^{\rho}{}_{\nu\mu}=\Gamma^{\rho}{}_{\mu\nu}-\Gamma^{\rho}{}_{\nu\mu}}$ and $\displaystyle{{f^c}{}_{ab} = {h_a}^\mu {h_b}^\nu \left( {{\partial _\nu }{h^c}_\mu  - {\partial _\mu }{h^c}_\nu } \right)}$ are the structure coefficients of tetrad basis satisfying $\displaystyle{\left[h_a,h_b\right]={f^c}{}_{ab}h_c}$.
\par The Fock-Ivanenko covariant derivative contains both torsion field part and pure gravity part. To explore the effect of gravity and torsion separately, it would be better to decompose the Fock-Ivanenko covariant derivative into $\mathcal{D}_\mu=\tilde{\mathcal{D}}_{\mu}-i K_{\mu}$ where $\displaystyle{\tilde{\mathcal{D}_\mu}=\partial_\mu-\frac{i}{2}\tilde {A^{ab}}_{\mu}S_{ab}}$ is the covariant derivative for the torsionless Levi-Civita connection and the effects of torsion are fully contained in the contortion part $\displaystyle{K_\mu=\frac{1}{2}{K^{ab}}_{\mu}S_{ab}}$. In the minimally coupling case, the Dirac action \eqref{original_S_D} can be reduced to 
\begin{equation}\label{MinCoup_S_D}
 S_D = \int {{d^4}x} \sqrt { - g} \bar \psi \left[ {i\left( {{\gamma ^\mu }{{\tilde {\mathcal{D}}}_\mu }\psi  - \frac{3}{4}i{\gamma ^a}{\mathcal{A}_a}{\gamma _5}\psi } \right) - m\psi } \right]
\end{equation}
where  $\gamma_5=i\gamma^0\gamma^1\gamma^2\gamma^3$ and $\displaystyle \mathcal{A}^a=\frac{1}{6}\varepsilon^{abcd}T_{bcd}$ is the axial vector part of torsion tensor\cite{Shapiro:1998zh}. The vector part $\displaystyle \mathcal{V}_a={T^b}_{ba}$ does not couple with the spinor fields in the minimally coupling case. However, the non-minimal coupling is universal because the renormalization counter terms of the coupling between matter field and torsion can generate the non-minimal coupling even the original coupling is the minimal one at tree level\cite{Shapiro:1998zh,shapiro2002physical,BUCHBINDER1985263}. Under the constraints of covariance, locality, dimension and parity preserving, the action of fermion in the curved spacetime background with torsion is
\begin{equation}\label{NonMinCoup_S_D}
S_D = \int {{d^4}x} \sqrt { - g} \bar \psi \left[ {i\left( {{\gamma ^\mu }{{\tilde {\mathcal{D}}}_\mu }\psi  + {i\eta _1}{\gamma ^a}{\mathcal{V}_a}\psi  - \frac{3}{4}i{\eta _2}{\gamma ^a}{\mathcal{A}_a}{\gamma _5}\psi } \right) - m\psi } \right]\;,
\end{equation}
where the possible allowed non-minimal terms reduce to the minimal coupling case when $\eta_1=0$ and $\eta_2=1$\cite{shapiro2002physical}.
\par The equation of motion(EoM)  of spinor field correspond to the action \eqref{NonMinCoup_S_D} can be easily read as
\begin{equation}\label{genEoM}
{i\left( {{\gamma ^\mu }{{\tilde {\mathcal{D}}}_\mu }\psi  + {i\eta _1}{\gamma ^a}{\mathcal{V}_a}\psi  - \frac{3}{4}i{\eta _2}{\gamma ^a}{\mathcal{A}_a}{\gamma _5}\psi } \right) - m\psi }=0.
\end{equation}
The equation \eqref{genEoM} is hard to solve for general torsion evolution. Fortunately, the solutions of spinor field for some special spacetime without torsion have been found.\cite{collas2019dirac} Moreover, since general relativity has passed almost all the observational examination\cite{Will:2005yc,Asmodelle:2017sxn,Crispino:2019yew}, the effects of torsion field should be relatively small compared to that of spacetime metric. Thus, we may take the torsion terms as perturbation to a torsion free theory. The torsion free Dirac action is 
\begin{equation}\label{S_0}
{S_0} = \int {{d^4}x} \sqrt { - g} \bar \psi \left[ {i{\gamma ^\mu }{\tilde {\mathcal{D}}_\mu }\psi  - m\psi } \right]
\end{equation}
with the corresponding equation of motion is
\begin{equation}\label{Eq_0}
{i{\gamma ^\mu }{\tilde {\mathcal{D}}_\mu }\psi  - m\psi } =0.
\end{equation}
The perturbation terms in the action can thus be read as
\begin{equation}\label{S_i}
{S_{I}} = \int {{d^4}x} \sqrt { - g} \bar \psi \left[{-\eta _1}{\gamma ^a}{\mathcal{V}_a}\psi  + \frac{3}{4}{\eta _2}{\gamma ^a}{\mathcal{A}_a}{\gamma _5}\psi  \right]= \int {{d^4}x}\mathcal{L}_I
\end{equation}
Since there is no time derivative terms of spinor field in the interacting Lagrangian given by \eqref{S_i}, the interact Hamiltonian density has the property
\begin{equation}
\int d^4x \mathcal{H}_{I}\left(x\right)=-\int d^4x \mathcal{L}_{I}\left(x\right)=-S_{I}.
\end{equation}
The S-matrix can then be calculated by 
\begin{equation}\label{total_Special_S-matrix}
 \textbf{S}=\textbf{T}\{e^{iS_I}\}\;.
\end{equation}

\subsection{Spinor fields in spatial-flat FRW spacetime}
\par Since we take torsion as perturbation to the torsion free theory in a fixed spacetime background, the spatial-flat FRW spacetime in our cases, we need to start from the solution of eq \eqref{Eq_0} in spatial-flat FRW background. The spatial-flat FRW metric is 
\begin{equation}\label{FRW_metric}
d{s^2} = d{t^2} - {a^2}\left( t \right)\left( {d{x^2} + d{y^2} + d{z^2}} \right)
\end{equation}
Substituting the metric \eqref{FRW_metric} into equation \eqref{Eq_0}, we can get the equation of motion for spinor field in FRW spacetime,  
\begin{equation}\label{Eq_FRW}
i\left( {{\gamma ^0}{\partial _t}\psi  + {a^{ - 1}}\vec \gamma  \cdot \vec \nabla \psi  + \frac{3}{2}\frac{{\dot a}}{a}{\gamma ^0}\psi } \right) - m\psi  = 0\;,
\end{equation}
where $\vec \gamma  \cdot \vec \nabla$ is just $\gamma^i \partial_i$. The general solution of Eq.\eqref{Eq_FRW} has the form\cite{PhysRevD.3.346} 
\begin{equation}\label{Gen_FRW_Psi}
{\psi _D} =\frac{1}{{(2\pi )}^3} \sum\limits_s {\int {{d^3}{{\vec p}_C}} } \frac{1}{{{\sqrt {2{E_{{{\vec p}_C}a}}{a^3}}} }}\left[ {{A_s}({{\vec p}_C},t){U_s}({{\vec p}_C},t){e^{ - i\Omega \left( {{{\vec p}_C}} \right)}} + {B^\dag }_s({{\vec p}_C},t){V_s}({{\vec p}_C},t){e^{i\Omega \left( {{{\vec p}_C}} \right)}}} \right]
\end{equation}
where phase factor of the exponential is $\displaystyle {\Omega \left( {{{\vec p}_C}} \right)}=\int_{ - \infty }^t d t'{E_{{{\vec p}_{C}}a}}\left( {t'} \right) - {{\vec p}_C} \cdot \vec x$ and ${\vec p}_C$ is the comoving coordinate 3-momentum rather than a physical one. The relation between the comoving 3-momentum and the physical one is $\displaystyle{\vec p_P} = \frac{{{{\vec p}_C}}}{{a\left( t \right)}}$ so that the energy $E_{{{\vec p}_{C}}a}$ is given by $\displaystyle E_{{{\vec p}_{C}}a}={\left( {\frac{{{{\left| {{{\vec p}_C}} \right|}^2}}}{{a{{\left( t \right)}^2}}} + {m^2}} \right)^{1/2}}$. The time-dependent four-component spinors ${{U_s}({{\vec p}_C},t)}$ and ${{V_s}({{\vec p}_C},t)}$ are ${U_s}({{\vec p}_C},t)={u_s}\left( {{{\vec p}_C}/a} \right)$ and ${V_s}({{\vec p}_C},t)={v_s}\left( {{{\vec p}_C}/a} \right)$ which satisfy $\displaystyle \left(\gamma^ap_a-m\right)u\left(\vec{p}\right)=0$ and  $\displaystyle \left(\gamma^ap_a+m\right)v\left(\vec{p}\right)=0$ with normalization $\bar u_r\left(p\right)u_s\left(p\right)=2m\delta_{rs}$ and $\bar v_r\left(p\right)v_s\left(p\right)=-2m\delta_{rs}$, while the time-dependent annihilation operators ${A_s}({{\vec p}},t)$ and ${B_s}({{\vec p}},t)$ for particles and anti-particles with spin $s$ and momentum $\vec{p}$ are the time-dependent (as well as $ \lvert \vec{p} \rvert$ dependent) complex linear combination of ${{A_s}({\vec p})}$ and ${{B_{-s}}^\dag({-{\vec p}})}$ and ${{B_s}({\vec p})}$ and ${{A_{-s}}^\dag({-{\vec p}})}$, i.e.
\begin{equation}
{A_s}(\vec p,t) = {D_{1,1}}\left( {\left| {\vec p} \right|,t} \right){A_s}(\vec p) + {D_{1, - 1}}\left( {\left| {\vec p} \right|,t} \right){B_{ - s}}^\dag ( - \vec p)
\end{equation}
and
\begin{equation}
{B_s}(\vec p,t) = {D_{ - 1, - 1}}^*\left( {\left| {\vec p} \right|,t} \right){B_s}(\vec p) + {D_{ - 1,1}}^*\left( {\left| {\vec p} \right|,t} \right){A_{ - s}}^\dag ( - \vec p) \;,
\end{equation}
with the non-vanishing anti-commutators of the operators $A,B$,
\begin{equation}
\left\{ {{A_r}\left( {{{\vec p}}} \right),{A^\dag }_s\left( {{{\vec q}}} \right)} \right\} = \left\{ {{B_r}\left( {{{\vec p}}} \right),{B^\dag }_s\left( {{{\vec q}}} \right)} \right\} = {\left( {2\pi } \right)^3}{\delta ^3}\left( {{{\vec p}} - {{\vec q}}} \right){\delta _{rs}}\;.
\end{equation}
The time evolution of complex linear combination factor $D_{a,b}\left(a,b=1,-1\right)$ is related to a factor $\displaystyle S\left(\left| {\vec p}_C \right|,t\right)= \dot a a^{-2}\left[2{E_{{{\vec p}_C}a}}^2\right]^{-1}m\left| {\vec p}_C \right|$. The evolution of $D_{a,b}$ leads to the evolution of average number of particles $\left\langle {{N_{\vec p}}\left( t \right)} \right\rangle $. The expanding rate of the universe $\dot a$ as well as the masses of neutrinos are small so that the factor  $\displaystyle S\left(\left| {\vec p}_C \right|,t\right)$ can be neglected. In this case, $D_{a,b}\left({\left| {\vec p} \right|,t}\right)$ will remain to be a constant,  i.e. $D_{a,b}\left({\left| {\vec p} \right|,t}\right)=\delta_{ab}$\cite{PhysRevD.3.346}. Hence, the Dirac spinor field in spatial-flat FRW spacetime can be written as
\begin{equation}\label{Lim_FRW_Psi}
{\psi _D} =\frac{1}{{(2\pi )}^3} \sum\limits_s {\int {{d^3}} } {{\vec p}_C}\frac{1}{{{\sqrt {2{E_{{{\vec p}_C}a}}{a^3}} }}}\left[ {{A_s}({{\vec p}_C}){u_s}\left( {{{\vec p}_C}/a} \right){e^{ - i\Omega \left( {{{\vec p}_C}} \right)}} + {B^\dag }_s({{\vec p}_C}){v_s}\left( {{{\vec p}_C}/a} \right){e^{  i\Omega \left( {{{\vec p}_C}} \right)}}} \right].
\end{equation}
\par To have well defined one-particle states, we may assume the scale factor $a\left(t\right)$ varies sufficiently smooth and approaches constant values $a_i$ and $a_f$ sufficiently fast as $t\to-\infty$ and $t\to+\infty$ respectively\cite{parker2009quantum}. The one-particle states with momentum $\displaystyle\vec{p}_P=\frac{\vec{p}_C}{a\left(t\right)}$ are created via creation operator with the comoving momentum ${A_s}^\dag({\vec p}_C)$\cite{parker2009quantum}. Moreover, a convenient Lorentz-invariant normalization in a finite box  $\left\langle\mathbf{p}_1 \mid \mathbf{p}_2\right\rangle^{(R)}=2 E_{\mathbf{p}_1} V \delta_{\mathbf{p}_1, \mathbf{p}_2}$\cite{maggiore2005modern} can be employed. In infinite-volume limit, we take $V\to\left(2\pi\right)^3\delta^3(0)$, hence at any fixed time $t$, the normalization should be given via
\begin{equation}
\left\langle\mathbf{p}_1 \mid \mathbf{p}_2\right\rangle=2E_{\mathbf{p}_1}\left(2\pi\right)^3\delta^3\left(\mathbf{p_1}-\mathbf{p_2}\right)=2E_{\mathbf{p}_1}a\left(t\right)^3\left(2\pi\right)^3\delta^3\left(\mathbf{p_1}_C-\mathbf{p_2}_C\right)
\end{equation}
Thus, the one-particle state at the fixed time $t$ should be given by $\displaystyle\left| {{{\vec p}_P},s} \right\rangle  = \left| {\frac{{{{\vec p}_C}}}{a},s} \right\rangle  = \sqrt {2{E_{\vec pa}}{a^3}} {A_s}^\dag \left( {{{\vec p}_C}} \right)\left| 0 \right\rangle $. If we set the initial state as one particle with momentum $\vec{k}_P$ and spin $s$ and the final state is a particle with momentum $\vec{k}^\prime_P$ and spin $r$, the initial and final state can be written as
\begin{equation}\label{i_state}
  \left| i \right\rangle  = \left| {{{\vec k}_P},s} \right\rangle  = \left| {\frac{{{{\vec k}_C}}}{{{a_i}}},s} \right\rangle  = \sqrt {2{E_{\vec k{a_i}}}{a_i}^3} {A_s}^\dag \left( {{{\vec k}_C}} \right)\left| 0 \right\rangle 
\end{equation}
and 
\begin{equation}\label{f_state}
\left| f \right\rangle  = \left| {\vec k{'_P},r} \right\rangle  = \left| {\frac{{\vec k{'_C}}}{{{a_f}}},r} \right\rangle  = \sqrt {2{E_{\vec k'{a_f}}}{a_f}^3} {A_r}^\dag \left( {\vec k{'_C}} \right)\left| 0 \right\rangle  .  
\end{equation}
\subsection{Scattering amplitude calculation}
\par Now we are ready to calculate the scattering amplitude. Since the torsion field in \eqref{S_i} is relatively small, it is convenient to expand the S-matrix \eqref{total_Special_S-matrix} to the 1st order i.e.
\begin{equation}\label{1st_S_matrix}
\mathbf{S}\simeq1+i\mathbf{T}\left\{S_I\right\}
\end{equation}
and the S-matrix element can thus be calculate via
\begin{equation}\label{S_matrix_GenElem}
\begin{gathered}
  {\left[ {{\mathbf{S}}} \right]_{fi}} = \left\langle f \right.\left| 1 \right|\left. i \right\rangle  + i\left\langle f \right.\left| {\mathbf{T}\left[ {{S_{I}}} \right]} \right|\left. i \right\rangle  \hfill \\
\end{gathered}
\end{equation}
Substituting the initial and final state \eqref{i_state} and \eqref{f_state} and the interact action (\eqref{S_i}) as well as the Dirac spinor field \eqref{Lim_FRW_Psi} into \eqref{S_matrix_GenElem}, the first term of \eqref{S_matrix_GenElem} is 
\begin{equation}
\left\langle f \right.\left| 1 \right|\left. i \right\rangle=\alpha {\delta ^3}\left( {{{\vec k}_C} - \vec k{'_C}} \right){\delta _{rs}}
\end{equation}
where $\displaystyle\alpha=2{\left( {2\pi } \right)^3}\sqrt {{E_{\vec k{a_i}}}{a_i}^3{E_{\vec k{a_f}}}{a_f}^3} $  is a factor related to normalization factor and the second term of \eqref{S_matrix_GenElem} is
\begin{equation}
i\left\langle f \right.\left| {\mathbf{T}\left[ {{S_{I_D}}} \right]} \right|\left. i \right\rangle=\frac{\alpha}{2{{\left( {2\pi } \right)}^3}}\int {{d^4}x} \frac{1}{{\sqrt {{E_{\vec k{'_C}a}}{E_{{{\vec k}_C}a}}} }}\left[ {{{{e^{i\left( {\int_{ - \infty }^t d t'\left( {{E_{\vec k{'_C}a}} - {E_{{{\vec k}_C}a}}} \right)\left( {t'} \right) - \left( {\vec k{'_C} - {{\vec k}_C}} \right) \cdot \vec x} \right)}}}}{{\bar u}_r}\left( {\vec k{'_C}/a} \right)X{u_s}\left( {{{\vec k}_C}/a} \right)} \right]
\end{equation}
where $\displaystyle X={ - {i\eta _1}{\gamma ^a}{\mathcal{V}_a} + \frac{3}{4}i{\eta _2}{\gamma ^a}{\mathcal{A}_a}{\gamma _5}}$ is the interaction vertex. The cosmic torsion fields satisfying cosmological principle can have only two independent non-zero components\cite{tsamparlis1979cosmological}, 
\begin{equation}
  {T_{ijk}} =  - F(t){\epsilon_{ijk}}
\end{equation}
and
\begin{equation}
{T^i}_{j0} = \mathcal{K}(t)\delta _j^i\;.
\end{equation}
The vector and axial vector parts of torsion thus have the form 
\begin{equation}
{\mathcal{V}_0} = 3\mathcal{K}\left( t \right)={\mathcal{V}_0}\left( t \right),{\mathcal{V}_i} = 0
\end{equation}
and 
\begin{equation}
{\mathcal{A}^0} =  - F\left( t \right),{\mathcal{A}^i} = 0.
\end{equation}
Because the 0th components of vector and axial vector parts of torsion which are the only non-zero one  are time dependent and space independent, the S-matrix element \eqref{S_matrix_GenElem} of Dirac spinor field can be simplified as
\begin{equation}\label{S_D_matrix_Elem}
{\left[ {{\mathbf{S}_D}} \right]_{fi}} = \alpha {\delta ^3}\left( {{{\vec k}_C} - \vec k{'_C}} \right)\left[ {{\delta _{rs}} - {i\eta _1}\int {dt} {\mathcal{V}_0}\left( t \right){\delta _{rs}} + \int {dt} {{\bar u}_r}\left( {{{\vec k}_C}/a} \right)\left[ {i{\eta _2}\frac{{3{\mathcal{A}_0}{\gamma ^0}{\gamma _5}}}{{8{E_{{{\vec k}_C}a}}}}} \right]{u_s}\left( {{{\vec k}_C}/a} \right)} \right]\;.
\end{equation}
The scattering amplitude is proportional to ${\delta ^3}\left( {{{\vec k}_C} - \vec k{'_C}} \right)$, which means that the particle will keep its comoving momentum after scattering so that the scattering is just a redshift to a particle. The $\vec{k}_C$ denpendence in the last term of the scattering amplitude implies the interaction rate is different for different initial momentum which may cause the change of temperature spectrum after scattering via torsion which will be discussed in the next section. 
\par Now we pay attention on the Majorana case. The $\psi_M$ and $\bar \psi_M$ are not independent for Majorana spinor field, i.e. $\displaystyle{{\bar \psi }_M}={\psi _M}^T\mathcal{C}$. The action of Majorana spinor is\cite{srednicki2007quantum} 
\begin{equation}
S_M=\frac{1}{2} \int {{d^4}x} \sqrt { - g} \bar \psi_M \left[ {i\left( {{\gamma ^\mu }{{\tilde {\mathcal{D}}}_\mu }\psi_M  + {i\eta _1}{\gamma ^a}{\mathcal{V}_a}\psi_M  - \frac{3}{4}i{\eta _2}{\gamma ^a}{\mathcal{A}_a}{\gamma _5}\psi_M } \right) - m\psi_M } \right]
\end{equation}
which is one-half of the Dirac one formally where $\psi_M$ is expanded by
\begin{equation}\label{Lim_FRW_Psi_M}
\psi_M= \frac{1}{{{(2\pi )}^3}}\sum\limits_s {\int {{d^3}} } {{\vec p}_C}\frac{1}{{\sqrt {2{E_{{{\vec p}_C}a}}{a^3}} }}\left[ {{A_s}({{\vec p}_C}){u_s}\left( {{{\vec p}_C}/a} \right){e^{ - i\Omega \left( {{{\vec p}_C}} \right)}} + {A^\dag }_s({{\vec p}_C}){v_s}\left( {{{\vec p}_C}/a} \right){e^{  i\Omega \left( {{{\vec p}_C}} \right)}}} \right]
\end{equation}
rather than $\psi_D$ given in \eqref{Lim_FRW_Psi}. Using the action $S_M$ and the Majorana spinor $\psi_M$, we can calculate the scattering amplitude as what we have done for the Dirac one. The 1st order term is 
\begin{equation}
i\left\langle f \right.\left| {\mathbf{T}\left[ {{S_{I_M}}} \right]} \right|\left. i \right\rangle=\frac{1}{4}\alpha\delta^3\left( {{{\vec k}_C} - \vec k{'_C}} \right)\int {{d}t} \left[ {\frac{1}{{E_{{{\vec k}_C}a} }}\left({{\bar u}_r}\left( {\vec k{'_C}/a} \right)X{u_s}\left( {{{\vec k}_C}/a} \right) - {{\bar v}_s}\left( {{{\vec k}_C}/a} \right)X{v_r}\left( {{{\vec k'}_C}/a} \right)\right)} \right]
\end{equation}
Here we use the fact that vertex is only dependent on time $X=X\left(t\right)$. Using the relationship that $v = C{{\bar u}^T}$ and ${C^\dag } =  - i{\gamma ^2}^\dag {\gamma ^0} = i{\gamma ^2}{\gamma ^0} =  - C$ we have 
\begin{equation}\label{key}
	{{\bar v}_s}(k){\gamma ^a}{v_r}(k') = {u_s}{(k)^T}C{\gamma ^a}C{{\bar u}_r}^T(k') = {u_s}{(k)^T}{\gamma ^a}^T{{\bar u}_r}^T(k') = {{\bar u}_r}(k'){\gamma ^a}{u_s}(k)
\end{equation}
and
\begin{equation}
\begin{gathered}
   {{\bar v}_s}(k){\gamma ^a}{\gamma _5}{v_r}(k') = {u_s}{(k)^T}C{\gamma ^a}{\gamma _5}C{{\bar u}_r}^T(k') =  - {u_s}{(k)^T}C{\gamma ^a}CC{\gamma _5}C{{\bar u}_r}^T(k')\\
   = {u_s}{(k)^T}{\gamma ^a}^T{\gamma _5}^T{{\bar u}_r}^T(k') =  - {u_s}{(k)^T}{\gamma _5}^T{\gamma ^a}^T{{\bar u}_r}^T(k') =  - {{\bar u}_r}(k'){\gamma ^a}{\gamma _5}{u_s}(k) \;.\\ 
\end{gathered}
\end{equation}
Then we have
\begin{equation}
i\left\langle f \right.\left| {\mathbf{T}\left[ {{S_{I_M}}} \right]} \right|\left. i \right\rangle=\frac{1}{2}\alpha\delta^3\left( {{{\vec k}_C} - \vec k{'_C}} \right)\int {{d}t} \left[ {\frac{1}{{2E_{{{\vec k}_C}a} }}\left({{\bar u}_r}\left( {\vec k{'_C}/a} \right)X^\prime{u_s}\left( {{{\vec k}_C}/a} \right)\right)} \right]
\end{equation}
where $\displaystyle X^\prime= \frac{3}{4}i{\eta _2}{\gamma ^a}{\mathcal{A}_a}{\gamma _5}$. In Majorana case, the axial vector part of torsion contributes the same to the scattering amplitude as in the Dirac case while the vector torsion has no effects on the scattering amplitude which is different from the Dirac case. In fact, for interaction Hamiltonian density $\bar \psi\Gamma \psi$, the vertex for Dirac field scattering is $\Gamma$ while effective vertex for Majorana field scattering is $\displaystyle\Gamma^\prime=\frac{1}{2}\left(\Gamma+C\Gamma^TC^{-1}\right)$ since effective vertex for Majorana field scattering should keep invariant under charge-conjugation transformation $\Gamma^\prime=C{\Gamma^\prime}C^{-1}$\cite{DENNER1992467}. The factor $\gamma^a\gamma_5$ keeps invariant under charge-conjugation transformation, i.e. $\displaystyle C\left(\gamma^a\gamma_5\right)^{T}C^{-1}=\gamma^a\gamma_5$, while the factor $\gamma^a$ is not, i.e. $\displaystyle C\left(\gamma^a\right)^{T}C^{-1}=-\gamma^a$, so that the field coupled with $\gamma^a$ has no contribution to scattering amplitude which is the vector part of torsion in our case. The total Majorana scattering amplitude is
\begin{equation}\label{S_M_matrix_Elem}
{\left[ {{\mathbf{S}_M}} \right]_{fi}} = \alpha {\delta ^3}\left( {{{\vec k}_C} - \vec k{'_C}} \right)\left[ {{\delta _{rs}} + \frac{3i\eta_2}{8}\int {dt} {{\bar u}_r}\left( {{{\vec k}_C}/a} \right)\left[ {\frac{{{\mathcal{A}_0}{\gamma ^0}{\gamma _5}}}{{{E_{{{\vec k}_C}a}}}}} \right]{u_s}\left( {{{\vec k}_C}/a} \right)} \right]
\end{equation}
which  differs \eqref{S_D_matrix_Elem} by the vector part of torsion.
\section{Shift of energy distribution}
\par As we mentioned before, the scattering amplitude is $\vec{k}_C$ dependent which will cause the shift of the energy distribution after scattering. Notice that the scattering amplitude \eqref{S_D_matrix_Elem} and \eqref{S_M_matrix_Elem} can be generally written as 
\begin{equation}\label{S_matrix_GenElem}
{\left[ {{\mathbf{S}}} \right]_{fi}} = \alpha {\delta ^3}\left( {{{\vec k}_C} - \vec k{'_C}} \right)M_{fi}
\end{equation}
The factor $\alpha$ is related to normalization. In torsion free spacetime, the scattering amplitude is ${\left[ {{\mathbf{S}}} \right]_{fi}} = \alpha {\delta ^3}\left( {{{\vec k}_C} - \vec k{'_C}} \right)$. The effect of torsion to the rate of redshift is included in the factor $M_{fi}$. The scattering rate from initial state $\left| i \right\rangle$ to the final state $\left| i \right\rangle$ $W_{fi}$ is proportional to $\left|\left[\mathbf{S}\right]_{fi}\right|^2$
\begin{equation}\label{GenW}
W_{fi}\propto\alpha^2\left(\delta^3\left(\vec{k}^\prime_C-\vec{k}_C\right)\right)^2\left|M_{fi}\right|^2
\end{equation}
Experimentally, it is more common that the initial spin is unknown to us and the detector receives all final spin configurations. Thus, to compare with experiment, we should sum all the final spin configurations and average the initial ones for $W_{fi}$. We define
\begin{equation}
\overline{W}_{fi}=\frac{1}{2}\sum_{\text {initial spins }} \sum_{\text {final spins }}W_{fi}\propto\alpha^2\left(\delta^3\left(\vec{k}^\prime_C-\vec{k}_C\right)\right)^2\overline{\left|M_{fi}\right|^2}
\end{equation}
where  
\begin{equation}
\overline{\left|M_{fi}\right|^2}=\frac{1}{2}\sum_{\text {initial spins }} \sum_{\text {final spins }}\left|M_{fi}\right|^2
\end{equation}
If the initial energy distribution is $I_{i}\left(E_i,T\right)$, the final energy distribution will be $I_{f}\left(E,T\right)=\bar{W}_{fi}I_{i}\left(E_{f}\left(E\right),T\right)$ where the $E_f\left(E_i\right)$ is the final energy dependency on the initial energy. Thus, if the final energy distribution in the torsion free background is $I_0\left(E,T\right)$, the final energy distribution in the spacetime with cosmic torsion will become $I_T\left(E,T\right)=\overline{\left|M_{fi}\right|^2}I_0\left(E,T\right)$. For Dirac field the scattering amplitude\eqref{S_D_matrix_Elem}  can be calculated as
\begin{equation}
\overline{\left|{M_D}_{fi}\right|^2}=1+\text{$\mathcal{V}$ terms}+\text{$\mathcal{A}$ terms}
\end{equation}
where  
\begin{equation}
\text{$\mathcal{V}$ terms}={{\eta _1}^2{{\left( {\int {dt} {\mathcal{V}_0}} \right)}^2} }
\end{equation}
and the $\mathcal{A}$ terms is 
\begin{equation}\label{A_terms}
\begin{gathered}
   - \frac{9}{{32}}{\eta _2}^2\int {dt'} \int {dt} \left( {\frac{1}{{{E_{{{\vec k}_C}a'}}\left[ {t'} \right]}}{\mathcal{A}_0}\left[ {t'} \right]\frac{1}{{{E_{{{\vec k}_C}a}}\left[ t \right]}}{\mathcal{A}^0}\left[ t \right]{k_{P;a'}}_c\left[ {t'} \right]{k_{P;a}}^c\left[ t \right] + \frac{1}{{{E_{{{\vec k}_C}a'}}\left[ {t'} \right]}}{\mathcal{A}_0}\left[ {t'} \right]\frac{1}{{{E_{{{\vec k}_C}a}}\left[ t \right]}}{\mathcal{A}^0}\left[ t \right]{m^2}} \right) \hfill \\
   + \frac{9}{{16}}{\eta _2}^2\int {dt'} \int {dt} \left( {{\mathcal{A}_0}\left[ {t'} \right]{\mathcal{A}_0}\left[ t \right]} \right). \hfill \\ 
\end{gathered} 
\end{equation}
For Majorana case, $\overline{\left|M_{fi}\right|^2}$ differs from the Dirac one by the term related to the vector part of torsion, i.e.
\begin{equation}
\overline{\left|{M_M}_{fi}\right|^2}=1+\text{$\mathcal{A}$ terms}
\end{equation}
\par Although neutrino is massless in Standard Model, the discovery of neutrino oscillations proved that neutrino is not exactly massless even though the mass of neutrino is very small. The upper limit on the absolute mass scale of neutrinos is 1.1 eV (90\% confidence level) according to recent experiment\cite{PhysRevLett.123.221802}. It is common that the energy of neutrino is much larger than its mass, i.e. $E\gg m$. In this case, ${E_{{{\vec k}_C}a}}$ can be expanded as 
\begin{equation}
{E_{{{\vec k}_C}a}} = {\left( {\frac{{{{\left| {{{\vec k}_C}} \right|}^2}}}{{a{{\left( t \right)}^2}}} + {m^2}} \right)^{1/2}} \simeq \frac{{\left| {{{\vec k}_C}} \right|}}{{a\left( t \right)}} + \frac{{{m^2}a\left( t \right)}}{{2\left| {{{\vec k}_C}} \right|}}
\end{equation}
and therefore the $\mathcal{A}$ terms can be simplified as
\begin{equation}
\text{$\mathcal{A}$ terms}\simeq\frac{9}{{32}}{\eta _2}^2 \left({2{\mathcal{A}_0}\left[ {t'} \right]{\mathcal{A}_0}\left[ t \right] - \frac{{{m^2}}}{{2{{{E_f}^2 {a_f}^2}}}}{{\left( {a\left( t \right) + a\left( {t'} \right)} \right)}^2}{\mathcal{A}_0}\left[ {t'} \right]{\mathcal{A}^0}\left[ t \right]}\right)\;.
\end{equation}
\par The dependence of $\overline{\left|{M}_{fi}\right|^2}$ on $E_f$ is also dependent on the mass $m$ and the axial vector part of torsion $\mathcal{A}$, which means if the particle mass is zero or there is no axial vector part of torsion, the interaction rates for particles with different energy are same which will cause that the final energy distribution scattering by torsion is the same as that given by torsion free case. In non-minimum coupling case that $\eta_1\neq 0$, the existence of the vector part of torsion will result in the same ${\left| {{{\vec k}_C}} \right|}^2$ dependence term in $\overline{\left|{M}_{fi}\right|^2}$ in both Dirac and Majorana case on the energy distribution. That is, even if the  ${\left| {{{\vec k}_C}} \right|}^2$ dependence term that can takes effect on the final energy distribution is the same for both types of spinor field, the final energy distribution for the two case are different for the same final energy distribution in torsion free case.
\par If the final energy distribution in torsion free case is one given by Planck formula for black body radiation,
\begin{equation}
{I_0}\left( {E,T} \right) = \left( {\frac{{{2E^3}}}{{{{\left( {2\pi } \right)}^3}}}} \right)\frac{1}{{{e^{\frac{E}{{{k_B}T}}}} - 1}}.
\end{equation}
The distribution reaches its peak at ${E_0}_{Max}$ given by
\begin{equation}
\frac{d{I_0}\left( {E,T} \right)}{dE}|_{E={E_0}_{Max}}=0
\end{equation}
which can be written as 
\begin{equation}\label{E0Max_Eq}
 \left( {3 - x} \right){e^x} = 3,
\end{equation}
where $x$ is defined as $\displaystyle x=\frac{E}{k_B T}$. The equation \eqref{E0Max_Eq} has a positive solution $x_0$. Therefore, the ${E_0}_{Max}$ can be given as
\begin{equation}
{E_0}_{Max}={x_0}{k_B}T
\end{equation}
which is known as the Wien's displacement law. 
\par Then we turn on the torsion field. In the Majorana case the distribution will shift to $I_M\left(E,T\right)=\overline{\left|{{M}_M}_{fi}\right|^2}I_0\left(E,T\right)$ whose peak is arrived at ${E_M}_{Max}$ given by $\displaystyle \frac{dI}{dE}=0$ which can be derived as 
\begin{equation}\label{EMMax_Eq}
\left( {3 - x} \right){e^x} = 3 + \zeta \left( {x,T} \right)\left( {{e^x} - 1 - x{e^x}} \right)
\end{equation}
where the shift factor
\begin{equation}
	\zeta \left( {x,T} \right) = \frac{\xi }{{\left( {1 + \chi } \right){{\left( {{k_B}T} \right)}^2}{x^2}}}
\end{equation}
and the factor
\begin{equation}\label{key}
	\chi  = \frac{9}{{16}}{\eta _2}^2\int {dt'} \int {dt\left( {{\mathcal{A}_0}\left[ {t'} \right]{\mathcal{A}_0}\left[ t \right]} \right)} 
\end{equation}
and
\begin{equation}\label{key}
	 \xi  = \frac{{9{m^2}}}{{64{a_f}^2}}{\eta _2}^2\int {dt'} \int {dt\left( {{{\left( {a\left( t \right) + a\left( {t'} \right)} \right)}^2}{\mathcal{A}_0}\left[ {t'} \right]{\mathcal{A}^0}\left[ t \right]} \right)} \;.
\end{equation}
The equation \eqref{EMMax_Eq} is hard to solve analytically, and the solution $x_M$ is dependent on the temperature rather than a constant like $x_0$. However, it can be noticed that the left-hand side of equation \eqref{EMMax_Eq} as well as the first term 3 in the right-hand side of \eqref{EMMax_Eq} is just the equation \eqref{E0Max_Eq} whose solution is known as the constant $x_0$. In the theory that the cosmic torsion plays the role of part of the dark energy, the cosmic torsion is in the order of Hubble parameter $H$\cite{Shen:2018elj,Zhai:2019std}. If the scale factor $a\propto t^n$, the integral $\displaystyle\int_{t_i}^{t_f}H\left(t\right) dt \propto ln\left(\frac{t_f}{t_i}\right)H_0t_0$ where $H_0$ is Hubble constant if the final time is today $t_0$. Since $H_0t_0$ is in the order of 1, the the integral $\displaystyle\int_{t_i}^{t_f}H\left(t\right) dt$ is in the order of 1. Hence, the factor $\chi$ is in the order about 1 and the factor $\xi$ is in the order $m^2$ so that shift factor $\zeta \left( {x,T} \right)$ is in the order of $\displaystyle \frac{m^2}{E^2}$ which is much less than 1 in the most cases for neutrinos. Therefore, the second term of the left-hand side of equation \eqref{EMMax_Eq} is small. Thus, the difference between the solution of eq.\eqref{EMMax_Eq} $x_M$ and the solution of \eqref{E0Max_Eq} $x_0$ is expect to be small. We may define the difference as $\Delta_M=x_M-x_0$ and rewrite the equation \eqref{EMMax_Eq} with $x_0$ and $\Delta_M$ in the first-order of $\Delta_M$ as 
\begin{equation}\label{EMMax_Eq_1st}
 {\Delta _M}{e^{{x_0}}} + \zeta \left( {{x_0},T} \right)\left( {{e^{{x_0}}} - 1 - {x_0}{e^{{x_0}}}} \right){\text{ = }}0
\end{equation}
and the $\Delta_M$ can be easily solved as 
\begin{equation}
{\Delta _M}= \frac{2}{3}\zeta \left( {{x_0},T} \right) =   \frac{{2\xi }}{{3\left( {1 + \chi } \right){{\left( {{k_B}T} \right)}^2}{x_0}}}.
\end{equation}
Thus the final Majorana energy distribution reaches it peak at
\begin{equation}\label{E_M_Max}
{E_M}_{Max} = {x_M}{k_B}T \simeq {x_0}{k_B}T + \frac{{2\xi }}{{3\left( {1 + \chi } \right)\left( {{k_B}T} \right){x_0}}}
\end{equation} 
and similarly the result of the Dirac one is 
\begin{equation}\label{E_D_Max}
{E_D}_{Max} \simeq x_0{k_B}T + \frac{{2\xi }}{{3\left( {1 + \chi  + V} \right)\left( {{k_B}T} \right){x_0}}}
\end{equation}
where
\begin{equation}\label{V}
 V={\eta _1}^2{\left( {\int {dt} {\mathcal{V}_0}} \right)^2}.
\end{equation}
  It can be seen that the shift of Dirac energy distribution and the Majorana energy distribution is different in the non-minimal coupling with $\eta_1$ and $\mathcal{V}_0$ is not zero.The energy shift is proportional to the order $\displaystyle \frac{m^2}{E^2}$. The energy distribution will not shift when $m=0$ because in such a case the interaction rate for all the energy is the same so that it will cause the same distribution after normalization. If we stick to Wien's displacement law to give the temperature, the effective temperature we detect will be 
\begin{equation}
{T_D}_{\text{eff}} = T\left( {1 + \frac{{2\xi }}{{3\left( {1 + \chi  + V} \right){{\left( {{k_B}T} \right)}^2}{x_0}^2}}} \right)
\end{equation}
for Dirac scattering and 
\begin{equation}
{T_M}_{\text{eff}} = T\left( {1 + \frac{{2\xi }}{{3\left( {1 + \chi } \right){{\left( {{k_B}T} \right)}^2}{x_0}^2}}} \right)
\end{equation}
for Majorana scattering. The temperature shift is also proportional to the order $\displaystyle \frac{m^2}{E^2}$.

\section{Conclusion and Discussion}
\par We find that the scattering amplitudes of the Dirac and Majorana fermions by torsion in spatial-flat FRW spacetime background are differed by the vector part of torsion field in non-minimal coupling case. The axial vector part of cosmic torsion coupled with mass in the interaction rate will cause a small shift of the energy distribution and the effective temperature derived from Wien's displacement law shift in the order $\displaystyle \frac{m^2}{E^2}$ if $m\ll E$ but $m\neq 0$. The shift of energy  distributions and effective temperature for Dirac and Majorana  spinors are different due to the difference of the scattering amplitude by the vector part of torsion.
\par The temperature of cosmic neutrino background (CNB) is given via $\displaystyle T_\nu=\left(\frac{4}{11}\right)^{1/3}T_{\gamma}\approx 1.95K$ with the corresponding energy $k_B T\sim 10^{-4}eV$\cite{weinberg2008cosmology}. However, the upper limit of neutrino mass is about 1.1eV\cite{PhysRevLett.123.221802}, which means the effect of static mass can not be neglect. The condition $m\ll E$ is no longer hold. Thus, the energy distribution shift is not the value given in \eqref{E_M_Max} and \eqref{E_D_Max} but that calculated  from the $\mathcal{A}$ terms given by \eqref{A_terms} combined with specific cosmological models with cosmic torsion. However, the basic qualitative conclusions that the term involving both axial vector part of cosmic torsion and the mass of neutrino will cause the energy distribution shift and the shift will be different for Dirac and Majorana scattering in the non-minimal coupling case that the vector part of cosmic torsion couples with the spinor field are still hold. 
\par The cosmic torsion of spatial-flat FRW spacetime takes simply effect on the rate of redshift in energy rather than the angle distribution of final state or leading to a change of the final state energy rather than redshift. This is due to the fact that the cosmic torsion in spatial-flat FRW spacetime is homogeneous isotropic and the scattering theory assume that the interaction is turned off in the initial and final state. The case of open and closed universe need to be discussed whether the interaction is spatially correlated so that it will caused the angle distribution of final state.

\section*{Acknowledgment}
This work is supported by the National Natural Science Foundation of China ( Grant Nos. 11775080, 11865016 ) and the Natural Science Foundation of Chongqing, China ( Grant No. CSTB2022NSCQ-MSX0351 ).
\clearpage
\bibliographystyle{unsrt}
\bibliography{ref}

\end{document}